\documentclass[10pt,conference]{IEEEtran}
\IEEEoverridecommandlockouts

\usepackage[british]{babel}
\usepackage[T1]{fontenc}
\usepackage[utf8x]{inputenc}
\usepackage{graphicx}
\usepackage{listings, xcolor}
\usepackage{fix-cm}\rmfamily  
\usepackage{paralist}
\usepackage{graphics}
\usepackage{amsmath}
\usepackage{booktabs}
\usepackage{siunitx}
\usepackage{tikz}
\usetikzlibrary{patterns}
\usepackage{multirow}
\usepackage[ruled,linesnumbered,noend]{algorithm2e}

\usepackage{lipsum}
\usepackage{gensymb}
\usepackage{colortbl}
\usepackage{tabularx}
\usepackage{pdflscape}
\usepackage{subcaption}
\usepackage{array}
\usepackage{enumitem}
\usepackage{listings}
\usepackage{ragged2e}

\usepackage{tikz}
\usetikzlibrary{shapes, arrows, positioning}

\tikzstyle{block} = [rectangle, draw, text width=0.95\columnwidth, align=left, minimum height=1cm]
\tikzstyle{arrow} = [thick,->,>=stealth]

\usepackage{pgfplots} 
\pgfplotsset{
	compat=1.9,
	compat/bar nodes=1.8,
}
\usepackage{threeparttable}
\usepackage{pgfplotstable}
\usetikzlibrary{patterns}
\usepackage{xcolor}
\usepackage[dvipsnames,table]{xcolor} 
\usepackage{amsfonts}
\usepackage{pifont}
\usepackage{xspace}
\usepackage{orcidlink}
\usepackage[numbers]{natbib}

\usepackage{todonotes}
\usepackage{comment}
\usepackage[multiple]{footmisc}
\usepackage{bigfoot}
\usepackage{rotating}
\usepackage{multicol}
\usepackage{caption}

\newcommand\goalstatement{\textit{The goal of this study is to aid software practitioners in taking informed actions as trust in the software supply chain evolves, through an interview study with 38 practitioners.}}

\newcounter{rqcounter}
\newcounter{subrqcounter}[rqcounter]
\newcommand\qt[1]{\textcolor{blue}{\textit{``#1''}}}
\newcommand\user[1]{{\footnotesize$\langle$#1$\rangle$}}

\newcounter{findingcounter}

\usepackage{titlesec}
\titlespacing*{\subsubsection}{0pt}{2pt plus 4pt minus 4pt}{2pt plus 0pt minus 2pt}

\newcommand{\newfinding}[2]{%
    \refstepcounter{findingcounter}%
    \label{#1}\vspace{.4em}%
    \textbf{\textcolor{Purple}{Finding \#\arabic{findingcounter}:}} \textcolor{Purple!95}{#2}%
}

\newcounter{trustcount}
\newcounter{distrustcount}

\newcommand\control[1]{\textcolor{OliveGreen}{\emph{#1}}}

\def\BibTeX{{\rm B\kern-.05em{\sc i\kern-.025em b}\kern-.08em
    T\kern-.1667em\lower.7ex\hbox{E}\kern-.125emX}}

\begin{document}

\title{The Rising Cost of Trust: Practitioners' Trust Signals, Controls, and Responses in the Software Supply Chain
}

\author{\IEEEauthorblockN{1\textsuperscript{st} Ranindya Paramitha, 2\textsuperscript{nd} Siri Paidipalli, and 3\textsuperscript{rd} Laurie Williams}
\IEEEauthorblockA{
\textit{North Carolina State University}\\
Raleigh, NC, USA \\
rparami,spaidip,lawilli3@ncsu.edu}
\and
\IEEEauthorblockN{4\textsuperscript{th} Christian K\"astner}
\IEEEauthorblockA{
\textit{Carnegie Mellon University}\\
Pittsburgh, PA, USA \\
kaestner@cs.cmu.edu}
}

\maketitle

\begin{abstract}
The software supply chain is becoming more complex, and AI is reshaping its threat landscape, e.g., raising concerns about the quality of AI-generated dependencies. 
Seen through the lens of \textit{trust}, the stakes of eroding trust in the software supply chain are high, yet we lack an empirical baseline on practitioners' trust.
\goalstatement\ 
We conducted semi-structured interviews with industry and open-source practitioners, focusing on their revealed preferences (the controls they adopted) rather than their stated attitudes, and analyzed the data using thematic analysis grounded in established trust concepts from the social sciences. 
We find that trust is eroding, which is becoming costly: aware practitioners are accumulating controls. 
To cope with the rising cost of trust, practitioners automate verification, delegate trust decisions to guardians, or consider exiting the software supply chain entirely.
Understanding software supply chain dynamics through the lens of trust provides the vocabulary and concepts (e.g., guardians of trust, system trust, signals) to shape future interventions for a well-functioning supply chain with appropriate levels of trust.
\end{abstract}


\section{Introduction}
\label{sec:introduction}
\textit{``Are practitioners losing trust in the software supply chain?''}
When we ask this question, we often hear ``well, yes, obviously'' as the answer.
Software supply chain attacks have become more prevalent, sophisticated, and visible in recent years ~(e.g., \cite{sonatype2026sscr, gokkaya2026software}), with incidents like the almost-successful XZ Utils attack~\cite{freund2024xz} now reported to nontechnical audiences~\cite{roose2024oneguy}.
At the same time, many admit that the situation is more nuanced with open-source communities, industry, researchers, and policymakers all taking significant steps toward improving software supply chain security~\cite{williams2025can}.
All the while, AI is reshaping the software supply chain with unclear consequences for trust.
AI enables more sophisticated social engineering attacks, facilitates the discovery and exploitation of vulnerabilities, and raises concerns about the low quality of generated dependencies~\cite{sonatype2026sscr, ji2024cybersecurity}, but AI is also used for defense~\cite{malkawi2026ai}.
The stakes of eroding trust are high: practitioners either have to bear the rising cost of verification or exit entirely (which AI may make increasingly feasible) and accept different risks.
Yet despite the high stakes, discussions around trust are mostly based on quick opinions and subjective impressions, without a shared understanding of what trust actually is. 
In this paper, we set out to understand how trust in the software supply chain is evolving.
Rather than relying on vibes and perceptions, we study the actions practitioners take, grounded in the rich trust concepts from social sciences.

\begin{figure}
    \centering
    \includegraphics[width=\columnwidth]{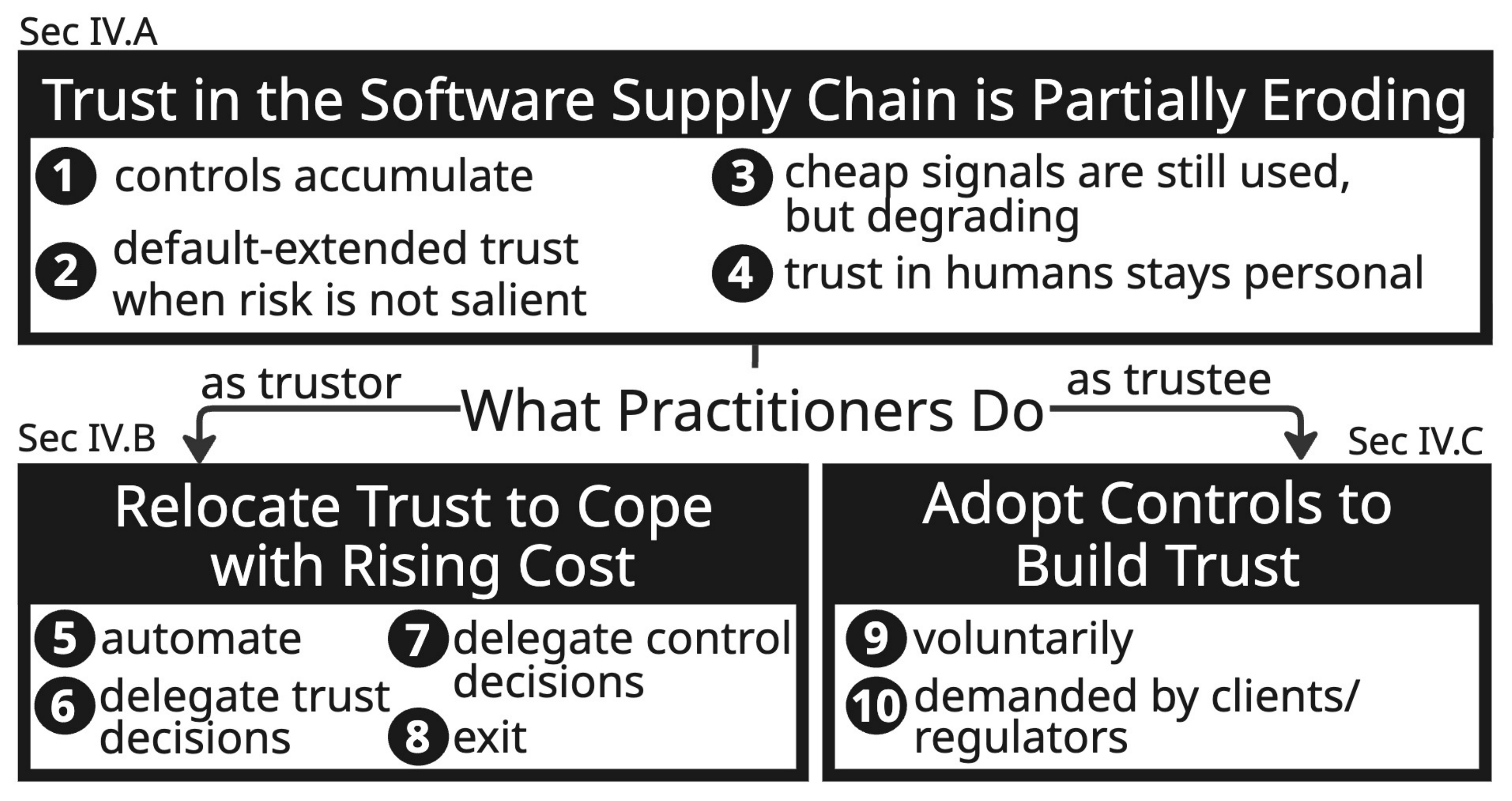}
    \caption{Overview: Trust in the Software Supply Chain is Partially Eroding, Practitioners Relocate and Build Trust.}
    \label{fig:summary}
\end{figure}



Trust is the willingness of one party to accept being vulnerable to the actions of another, based on positive expectations about the other's actions, without the ability to monitor or control them~\cite{mayer1995integrative}. Historically, open-source communities have been high-trust environments in which developers rely on others' code without rigorous checks or monitoring~\cite{raymond1999cathedral, de2010can}. Trust is described as the \textit{\textbf{lubricant} of social and economic systems}~\cite{arrow1974limits}, because it enables collaboration with low transaction costs (e.g., high-trust societies spend less on lawyers and security guards~\cite{fukuyama1995trust,putnam2000bowling}). The inverse, mistrust, requires extensive checking. \looseness=-1

In software engineering, trust in the software supply chain enables reuse of third-party dependencies at near-zero transaction costs (``the lubricant''). 
When trust erodes, practitioners take precautions, e.g., running extra tools, hardening builds, auditing dependencies, at great cost.
One practitioner in our interview expressed the tension: \qt{I see the necessity for [the controls].}, yet \qt{it does get frustrating when all of these controls are enforced.} 
Many companies each have their own internally audited versions of subsets of open-source ecosystems~\cite{Zahan2024-vz},
each separately paying for auditing (the equivalent of each hiring their own security guards). If trust erodes, we might undermine many of the benefits of open source through high transaction costs.

However, asking practitioners directly about their \textit{trust} in the software supply chain (e.g., on a rating scale) is unlikely to produce reliable data or deeper insights.
Intuitive interpretations of trust vary widely between people~\cite{hou2023systematic}, and answers may be tainted by various response biases~\cite{paulhus1991measurement}. 
Instead, using \textit{revealed preference theory}~\cite{samuelson1938, glaeser2000measuring}, we focus on the actions practitioners take (or do not take) and interpret these actions through the lens of trust.
That is, rather than relying on what practitioners \textbf{say} (stated preferences), we study what they \textbf{do} (revealed preferences).

\goalstatement\xspace
We performed a qualitative interview study with 38 practitioners to capture practitioners' revealed preferences, focused on one central research question:
%
\textit{\textbf{``How is practitioners' trust in the supply chain evolving?''}}
{
}




We find that developers' control actions accumulate and are rarely retired, indicating eroding trust in the supply chain as a system. Rather than abandoning trust, practitioners relocate it to keep the cost bearable: they automate checks and delegate to intermediaries who act as guardians of trust, pooling effort and balancing risk exposure against cost. 
Clients and regulators increasingly demand controls, underwriting reliance through formal requirements.

Overall, we find that developer actions and decision-making have evolved substantially, suggesting shifts in trust. 
We find emerging trust mechanisms predicted in the social science trust literature: intermediaries who serve as \textit{guardians of trust}, pooling efforts and balancing risk exposure against costs. 
Understanding these parallels provides the vocabulary and concepts for designing more targeted interventions.

We make the following contributions:
\begin{itemize}[leftmargin=*]
    \item A discussion of software supply chain security through the lens of trust, grounded in a deep intellectual history of trust in the social sciences.
    \item A study design to elicit a deeper understanding of developers' trust in the software supply chain using revealed preferences theory.
    \item Results on the evolution of trust in the software supply chain from an interview study with 38 practitioners. These results can also serve as a baseline for future studies exploring continued behavior trends in software supply chain security.
    \item Discussion about lessons that a long history of research on system trust can offer for software supply chain security.
\end{itemize}


\section{Trust in Software Supply Chain: Background, Framing, and Related Work}
\label{sec:background_relwork}

Software development scales through modularity and reuse. 
Information hiding, 
allows a developer to use a component through its interface without inspecting (or even understanding) its internals~\cite{parnas1972criteria}. 
A software supply chain extends that principle across organizational and authorship boundaries~\cite{decan2019empirical}: developers reuse functionality written by others (who also reuse others') without understanding it.
For example, writing React applications without knowing how React is implemented and what React reuses internally.
Yet, reuse without inspection requires trust that the component behaves as expected.
Modern software supply chains involve hundreds or thousands of transitive dependencies~\cite{soto2021comprehensive}, making trust essential: components must be relied on without first-hand verification.
This reliance is, at its core, a \textbf{relationship of trust} as studied across the social sciences.

\subsection{Trust and control}
The term ``trust'' appears frequently across software engineering and security literature, e.g.,~\cite{kula2015trusting,wermke2022committed}, though its meaning is rarely made explicit.
We adopt an explicit definition from social science: the willingness of one party to \textbf{accept being vulnerable to the actions of another}, based on positive expectations about that party's conduct and held irrespective of the trustor's ability to monitor or control them~\cite{mayer1995integrative}.
\textit{Trust is generally societally beneficial:} it allows cooperation with low transaction costs (the ``lubricant'')~\cite{arrow1974limits}, whereas low-trust environments pay for repetitive checking of every transaction~\cite{fukuyama1995trust,putnam2000bowling}.
Open-source software has traditionally been an instance of a high-trust regime: a global community produces components that others depend on, with minimal cost to adopt or check. 
Software supply chain attacks threaten to undermine the conditions under which that low-cost cooperation was possible.\looseness=-1

In the software supply chain setting, the object of trust is a sociotechnical arrangement: the maintainers of a dependency, the infrastructure through which it is built and distributed, the norms of the community that produces it, and the broader ecosystem in which all of these are embedded. 
This arrangement, where the object of trust is a configuration of people, infrastructure, and norms rather than a single party, is what Luhmann~\cite{luhmann1979trust} distinguishes from personal trust as \textit{system trust}.
System trust is a trust extended to an arrangement whose continued functioning makes individual interactions safe enough to act on without further verification. 
As a classic example, a customer could, in theory, study monetary policy before accepting banknotes, but usually does not because trusting the monetary system is cheaper than verifying it.
In adopting a dependency, a developer accepts vulnerability to the maintainer, the infrastructure, and community processes, implicitly judging the arrangement as a whole trustworthy enough to justify the exposure.\footnote{To avoid confusion between security vulnerabilities and the act of being vulnerable to  somebody's actions, we refer to the latter as risk exposure.} 
Note that trust may also be misplaced and naive, for example, underestimating the exposure through limited knowledge, inexperience, or inattention. In this paper, we make no judgment about what trust is appropriate.

The complement to trust is \textbf{control}: checking, monitoring, and verification. 
In practice, control is rarely that thorough; cheaper, weaker forms of verification that reduce rather than eliminate exposure. 
For example, most security actions developers take are partial: checking only some updates, treating a signal such as README quality as a proxy for code quality, running static analysis for suspicious patterns, or judging the maintainer rather than the code~\cite{pashchenko2020qualitative,hamer2025trusting,dabbish2012social}. 
Each such act of control signals that the developer is unwilling to be fully exposed to the dependency's risk, and the degree of control exercised reflects the extent to which trust has been withheld.\footnote{We treat the control actions as inversely reflecting trust, that is, as substitutes: a developer who verifies more is, to that degree, unwilling to rely without checking.
An established line of work instead treats trust and control as complements, where control structures build trust by lowering the cost of misplaced reliance~\cite{DasTeng1998,PoppoZenger2002}.
We read the two as operating at different levels: discrete, developer-initiated actions (e.g., auditing dependency updates) are substitutes in this narrow sense.
Shared institutional infrastructure (e.g., signing systems, registries) sustains system trust and is better read as a condition of the high-trust regime than as a per-developer indicator of withheld trust. 
The same infrastructure can, however, also be adopted as a costly trustworthiness signal to downstream users~\cite{Spence1973} rather than as a substitute for trust; motivation distinguishes the two roles.
}

As suggested by Boughton et al.~\cite{boughton2024decomposing}, we ground trust measurement in observable control actions rather than stated attitudes: controls are more directly observable than trust itself, yet support inferences about it.
We therefore do not ask practitioners about their \textit{perception} of trust (which is subjective and may not align with our conception of trust).
We instead ask about the control actions they take and how these have changed over time. 
Using revealed preference theory~\cite{samuelson1938}, we study \textit{revealed preferences} rather than \textit{stated preferences}, which are generally 
more reliable and 
less susceptible to response biases and misunderstandings~\cite{paulhus1991measurement}.

\subsection{{Signals and the Breakdown of Honest Signaling}}

Unless we see wide adoption of proof-carrying code~\cite{necula1997proof} or similar strategies, direct verification is impractical at scale.
Thus, practitioners rely on observable proxies for a component's trustworthiness, e.g., in adopting dependencies, developers infer quality and good intent from signals such as a project's star count, contributor activity, README quality, or the maintainer's affiliation with a well-known company~\cite{dabbish2012social,wermke2022committed,pashchenko2020qualitative, hamer2025trusting, bogart2016break, larios2020selecting, mujahid2023characteristics, qiu2019signals}.
\textit{Signaling theory}~\cite{Spence1973} provides a useful framing for why such proxies can be informative: a signal that carries information is a signal that is cheaper for a trustworthy participant to produce than for an untrustworthy one.
For example, having a long maintenance history
was historically cheaper for competent and engaged maintainers to produce than for negligent or absent maintainers.
The cost asymmetry allowed developers to read these signals as evidence of trustworthiness and decide to trust rather than fully verify the code~\cite{wermke2022committed}.\looseness=-1

Bad actors in the software supply chain try to game these signals. 
Attackers can build a plausible profile of a reputable developer (as happened in the XZ Utils attack~\cite{freund2024xz}) and buy stars~\cite{he2026six} to make projects appear popular. 
Increasingly, generative AI makes mimicry cheap, for example, by generating a project with a high-quality README and lowering issue-response times with automated responses.
If faking a signal becomes cheap, the signal ceases to be reliable (known as the ``breakdown of honest signaling'')~\cite{searcy2005evolution}, leading to a \textit{``market for lemons''} where customers are unable to distinguish high-quality from low-quality products~\cite{akerlof1970market}.
In the open-source context, developers may abandon easy-to-mimic signals or come to distrust all dependencies equally.

\subsection{{The Role of Attackers and the Shift to System Trust}}

While there are cases where maintainers deliberately attack their users (e.g., ``protestware''~\cite{finken2025characteristics}), typically the maintainer of a depended-upon dependency is not the source of harm but an additional target.
The traditional trust framework for interpersonal trust can partially accommodate this case by treating it as a question of the trustee's \textit{ability}~\cite{mayer1995integrative}.
However, framing the problem purely as a matter of maintainer ability blames individuals for failures in tightly coupled, complex systems~\cite{Perrow1984}, especially when attackers can invest substantial resources, as in the XZ Utils attack.
A developer who depends on a dependency is exposed not only to the maintainer's conduct but also to adversaries who target the maintainer, upstream dependencies, and the build and distribution infrastructure.\looseness=-1

We adopt \textit{system trust} rather than \textit{interpersonal trust}: we are not just interested in trust in individual maintainers, but in confidence in the software supply chain \textit{as a system} that makes such trusting individual maintainers or artifacts reasonable.
The attacker is not a party to any trust relationship but a degrader of trust-supporting properties, undermining signals, infrastructure reliability, and the assumption that well-placed trust will not be weaponized.
Control actions should therefore be read as practitioners' assessments of the supply chain as a whole, not of any individual maintainer.

\subsection{Related Empirical Studies on Trust in Supply Chains}
\label{sec:relwork_trust_sc}

There are four empirical studies that are relevant to ours.
Jansen et al.~\cite{jansen2021trustseco} interviewed 12 practitioners
and found that organizational factors are more important than technical factors.
Pashchenko et al.~\cite{pashchenko2020qualitative} interviewed 25 practitioners 
and found that security concerns play a secondary role to functional concerns in developers' decisions.
Sammak et al.~\cite{sammak2023developers} interviewed 18 practitioners to explore broader practices for securing the software supply chain. They found that developers care about software supply chain security, yet remain unaware of threats and mitigations.
Wermke et al.~\cite{wermke2022committed} conducted interviews with 27 practitioners, explored trust models involving contributors and trust incidents, and found no uniform security or trust practices across projects. 
Our work complements this literature by exploring the connection between what practitioners trust and what that trust leads them to do, within a rich social science context.
We also expand to consider who makes trust decisions and why in the software supply chain.
Note that we focus on practitioners' trust, 
not on the software supply chain's trustworthiness~\cite{bauer2019conceptualizing}. 


\section{Methodology}
\label{sec:methodology}
We conducted 38 semi-structured interviews with industry and open-source practitioners (December 2025--April 2026), discarding 2 additional interviews due to credibility concerns. 
We discussed practices for managing and securing the software supply chain, including dependency management, build systems, human factors, and processes. 
We designed the study deliberately to understand trust and its evolution, but did not use the term ``trust'' in any of our recruiting materials or questions.  
The study was approved by our institution's Institutional Review Board (IRB).
All participants provided informed consent; transcripts were anonymized, and data were collected and stored per IRB and GDPR guidelines. 

\subsection{Scope: Software Supply Chain Security}
\label{subsec:scope}
We scope our study to software supply chain security, i.e., the integrity of everything a software product incorporates from outside its own codebase~\cite{williamsDirection2025}, though practitioner discussions sometimes spill into broader security concerns.
Our scope covers the third-party and open-source dependencies a project depends on (directly and transitively), the infrastructure that builds and distributes it, and the people who develop and maintain both. 
In scope are the controls practitioners use to manage risk exposure:
Regarding dependencies, controls include \control{scanning for known vulnerabilities},
\control{vetting a dependency before adoption}, and \control{maintaining audited internal mirrors}. 
Regarding build and distribution, controls include \control{artifact signing and verification},
\control{reproducible builds}, \control{software bills of materials (SBOMs)}, and \control{CI/CD hardening}.
Regarding human factors, controls include \control{vetting maintainers}, 
\control{multi-party code review} to catch a malicious or compromised contributor, and \control{insider threat monitoring}. 
Note that with Luhmann~\cite{luhmann1979trust}'s concept of system trust, practitioners can extend trust beyond the three components of the software supply chain to the system as a whole.
Beyond scope are general security actions independent of any reuse relationship, e.g., securing first-party code, implementing network defenses, and encrypting data. 
The boundary is context-dependent: \control{vulnerability scanning} is in scope when applied to dependencies or build configuration, but not when applied only to first-party code.


\subsection{Study Design}
Our core design decision was to study trust in the software supply chain through practitioners' actions, that is, through their \textit{revealed} rather than their \textit{stated} preferences. 
Direct questions about trust suffer from construct validity issues because intuitive interpretations vary widely~\cite{McKnight01122001}.
The cost of this choice is that we only interpret trust from practitioners' practices rather than as a direct measure, an implication we return to in Section~\ref{sec:limitations}.
The benefit is that our questions rest on concepts participants share with us and with each other. 
We opted to conduct \textit{semi-structured interviews} with practitioners from both open source communities and industry, as many practices of interest leave little public trace, and public artifacts offer at most a partial window into industry practice.
Interviewing practitioners lets us elicit practices broadly and explore unanticipated directions in follow-up questions. 

Our framing of practices rather than opinions helps reduce (though not eliminate) some \textit{response biases}.
By anchoring questions on concrete practices and their timing, we reduce \textit{recall bias}: practitioners remember specific technical decisions, 
more reliably than they recall how their attitudes evolved over the years. 
We similarly reduce incentives for \textit{social desirability bias} by asking about changing practices rather than their personal stance. 
Also, \textit{hindsight bias} is a lesser concern here, as salient events such as the XZ Utils incident and, especially, documented changes to company policy give participants firm anchors for what changed and when.  

We structured our interviews (after intro and demographic questions, about 10 min) into separate phases exploring practices on dependencies (e.g., adopting/updating dependencies, about 15 min), build infrastructure (e.g., GitHub actions, about 15 min), and human factors (e.g., insider threats, external contributors, about 10 min), before closing with questions about general perceptions of software supply chain security (about 10 min). 
To anchor each phase concretely, we started by asking about the last time they did a practice (e.g., adopted a dependency, updated a dependency, added build workflows) and had them walk us through the steps involved, then asked more broadly about their practices. 
For each practice, we ask follow-up questions about why they adopted it, when they started using it, and whether they stopped practicing it. 
The guide was refined through 7 pilot interviews.
The interview guide is made available in our Zenodo artifact~\cite{zenodo} (preview only, will be published after acceptance).
We do not share the recordings and transcriptions to protect participants' privacy and comply with IRB requirements.

We conducted interviews over Zoom and scheduled each interview for 60 minutes.
We recorded the interviews (with the participants' consent) and used anonymized transcripts for analysis.
For 14 interviews in which some information was missed during the interview, we emailed participants clarification questions, quoting their answers regarding a control.
Of the 14 emails sent, 12 participants replied, and we added their responses to our transcripts for analysis.

\subsection{Recruitment and Participants}
We recruited 19 industry practitioners and 19 open-source practitioners. 
All recruited participants have experience managing at least one of the three risk exposures (dependencies, build infrastructure, or human factors), and most open-source developers contribute dependencies to the supply chain. 
We used a mix of recruitment strategies: 
    \textbf{(1)~Professional Connections:} we contacted our professional network, including industry researchers and CISOs.
    \textbf{(2)~Events and Conferences:} we pitched our research at local conferences and events [details omitted for anonymous review].
    \textbf{(3)~Social Media Platforms:} we posted on LinkedIn, Slack workspaces, developers' Discord servers.
    \textbf{(4)~Open Source:} 
    we analyzed public GitHub events over 2 months through the GitHub API to identify practitioners who had recently added dependencies or adopted new workflows in projects with at least 10 stars. We manually analyzed 669 projects that adopted a new dependency and 267 that extended their builds, filtered maintainers by IRB location criteria, and contacted 395 practitioners with public email addresses within 2 weeks of the relevant change.

Interested practitioners completed a screening form covering eligibility criteria (experience with dependencies, builds, or human factors; age $\geq$18; not located in IRB-identified countries of concern) and provided consent.
Our \textit{open source} recruitment strategy was the most effective for reaching open-source practitioners (20 responses, 5\% response rate).
We interviewed all 21 open-source practitioners (1 from \textit{events and conferences}), yet had to discard 2 due to credibility issues.
For industry participants, we reached 12 through \textit{professional connections}, 3 through \textit{events and conferences}, and 4 through \textit{social media platforms}. These channels reach different populations, and the resulting selection effects bound what our sample represents. 
Recruiting most industry participants (12 of 19) through our professional network favors practitioners already engaged with supply chain security. For the open-source channel, targeting maintainers who have just made a dependency or workflow change favors active over dormant maintainers.
We accepted these effects in exchange for reaching a specialist, hard-to-contact population, and discuss transferability in Section~\ref{sec:limitations}.


We list the demographics of our 38 participants in Table~\ref{tab:participants-short}. 
Most come from North America (19), with representation from Asia (11), Europe (7), and Africa (1). 
Roles range widely from freelancers to chief officers; 19 have more than 5 years of experience, and 19 have fewer than 5 years.
The median interview lasted 57 minutes and 39 seconds.

\begin{table}[]
    \centering
    \scriptsize
    \caption{Participant Demographics}
    \label{tab:participants-short}
    \resizebox{\columnwidth}{!}{
    \begin{threeparttable}
    \begin{tabular}{lllll}
         \hline
            \textbf{ID} & \textbf{Role$^{\dagger}$} & \textbf{Company Type} & \textbf{Exp.} & \textbf{Country} \\
        \hline
            \rowcolor{gray!20}
            P1 & Senior security eng. & Software vendor & 3-5 & NLD \\          \rowcolor{gray!20}
            P2 & Software eng. & Telecommunications & 3-5 & IDN \\
            \rowcolor{gray!20}
            P3 & Senior software eng. & Enterprise software & $>$10 & USA \\
            P4 & Director & Consulting & $>$10 & DEU \\
            P5 & AI eng. & Tech networking & 6-10 & USA \\
            P6 & Senior software eng. & Software vendor & $>$10 & USA \\
            \rowcolor{gray!20}
            P7 & ML eng. & AI  & 0-2 & IND \\
            \rowcolor{gray!20}
            P8 & Software eng. & Fintech & 3-5 & USA \\
            \rowcolor{gray!20}
            P9 & Business analyst & IT services/consulting & 0-2 & IND \\
            \rowcolor{gray!20}
            P10 & IT security & IT services/consulting & $>$10 & USA \\
            \rowcolor{gray!20}
            P11 & Executive & Cybersecurity & $>$10 & CAN \\
            P12 & Software eng. & AI  & 0-2 & IND \\
            \rowcolor{gray!20}
            P13 & Software eng. & Sustainability tech & 3-5 & CAN \\
            \rowcolor{gray!20}
            P14 & Software eng. & Business intelligence & 3-5 & USA \\
            P15 & Freelancer & AI  & 0-2 & GBR \\
            \rowcolor{gray!20}
            P16 & IT security & IT services/consulting & 3-5 & USA \\
            P17 & Student & Higher education  & 6-10 & SGP \\
            P18 & Student & Higher education  & 3-5 & KOR \\
            P19 & Executive & Consulting & $>$10 & USA \\
            \rowcolor{gray!20}
            P20 & Director & Healthcare products & $>$10 & GBR \\
            P21 & Executive & Data analytics & $>$10 & USA \\
            \rowcolor{gray!20}
            P22 & Executive & Fintech & 0-2 & USA \\
            \rowcolor{gray!20}
            P23 & Executive & Consulting & $>$10 & USA \\
            \rowcolor{gray!20}
            P24 & Software eng. & Banking \& financial & 3-5 & IND \\
            \rowcolor{gray!20}
            P25 & Software eng. & E-commerce & 3-5 & IND \\
            P26 & Senior software eng. & IT services & $>$10 & ROU \\
            P27 & Software eng. & Data analytics \& AI & 3-5 & NGA \\
            P28 & Senior software eng. & Fintech & 6-10 & UKR \\
            P29 & Executive & Compliance & $>$10 & USA \\
            P30 & Executive & AI research  & 6-10 & USA \\
            \rowcolor{gray!20}
            P31 & Software eng. & IT services & 3-5 & IND \\
            P32 & Executive & Software \& AI & 0-2 & USA \\
            P33 & Researcher & Higher education  & 6-10 & SGP \\
            P34 & Senior software eng. & IT services & $>$10 & USA \\
            \rowcolor{gray!20}
            P35 & Executive & Consulting & 0-2 & USA \\
            \rowcolor{gray!20}
            P36 & Senior devops & IT services & $>$10 & IND \\
            P37 & Student & Higher education & 3-5 & GBR \\
            P38 & Software eng. & Digital health tech & $>$10 & USA \\
        \hline
    \end{tabular}
            \begin{tablenotes}
            \scriptsize
                \item[] 
                Gray: Industry practitioner;
                Exp: Experience in years;
                $\dagger$ abstracted for privacy
        \end{tablenotes}
    \end{threeparttable}
    }
\end{table}



\subsection{Analysis}
\label{subsec:data_analysis}

\noindent We analyzed 928 pages of transcripts from 38 interviews using thematic analysis with iterative theme refinement~\cite{fereday2006demonstrating,braun2006using}.

\textbf{Coding:} We started with open coding phase (after the first 10 interviews, then updated as we conducted more interviews), 
focusing on controls related to supply chain security (see Scope, Sec.~\ref{subsec:scope}). 
Especially, we looked for (a) any controls mentioned (e.g., \control{vetting dependency}, \control{attestations}); (b) any reasons for or against adopting the controls (e.g., past attacks); (c) any mentions of changes in controls (e.g., X started recently, replacing Y); and (d) any reasons for changes (e.g., company growth).
We incrementally developed a codebook,
totaling 377 codes in the end.
Two authors performed this coding independently and then reached consensus on the codes. 
This coding also segmented each interview transcript into discrete chunks, one chunk per distinct control mentioned.
To analyze the data through the lens of trust theory, we then converged on a narrower deductive coding approach based on a 53-code codebook derived from the previous open coding.
Using the chunks from the previous coding, for each chunk, coders made a forced choice along each of five dimensions of the codebook: (a) who decided the control (trustors, e.g., participant, participant's institution); (b) who or what the trust is (not) placed in (common trustees, e.g., humans, tools, systems); (c) whether trust is imposed or extended implicitly; (d) why the control decision is made; and (e) how the control has changed.\looseness=-1

\textbf{Theme Development and Finding Synthesis:}
Throughout various stages of the coding process, all authors met regularly to discuss and track insights and themes on Miro~\cite{miro}.
Among others, we grouped controls and targets of trust they imply and iteratively identified the themes we report, such as distinctions between personal and system trust, delegation of trust decisions, and external forces that shape control adoptions.


\textbf{LLM-Assisted Reliability Improvement:}
We used a large language model (LLM) (claude-sonnet-4-6 under commercial terms that exclude submitted data from model training; with IRB approval) as an \textit{additional}, \textit{redundant} effort to catch nuances humans might have missed
in the final deductive coding.
We prompt the LLM with the codebook and the chunks of interview transcripts from the \textit{coding} phase (each on a different control) to identify coding dimensions.
One author manually reviewed the LLM output and updated our coding where relevant, and another author reviewed those changes again (both authors divided the work and swapped the primary/secondary reviewer roles for half the interviews). Note that the use of LLM in thematic analysis is controversial~\cite{jowsey2025we}, but we deliberately used it as an \textit{additional} mechanism to increase reliability, not as a mechanism to \textit{replace} the researchers' judgment.\looseness=-1


\subsection{Limitations}\label{sec:limitations}
We record limitations using established criteria for qualitative trustworthiness~\cite{lincoln1985naturalistic} so readers can calibrate their inferences.
Our sampling surfaces relevant patterns but is not designed for quantification: participants who agreed may hold stronger views or better practices than those who declined (self-selection).
English-only interviews from a limited set of countries further narrow the pool.
We therefore describe the range of practices and reasoning we observed when we reached saturation, with context for readers to judge transferability, rather than population distributions.
Recalled actions are imperfect: participants may omit information that reflects poorly on them (social desirability bias), or reconstruct reasons from hindsight (recall and causal attribution bias).
Our focus on concrete actions
limits but does not eliminate these biases.

Most importantly, inferring trust from control actions is a reading rather than a measurement, and another analyst might interpret the same action differently.
This limitation is inherent to interpretive research, whose value lies in a coherent and well-grounded account. 
We worked toward such an account through repeated iteration and consensus among four authors, and grounded our interpretation in close engagement with the trust literature.
We report our coding frame (see Zenodo~\cite{zenodo}) and findings transparently so readers can follow and contest our interpretation. 



\section{Results}
\label{sec:results}
\label{subsec:rq1}



We report results in three parts: (a) how trust is evolving and how this comes with increased costs for controls, (b) how practitioners adapt to cope with lower direct trust and partially relocate trust, and (c) how external forces like customer demands shape control adoption and system trust.

\subsection{Partially Eroding Trust in the Software Supply Chain}
\label{subsec:result_part1}

Most participants were clearly aware of the constant stream of reported supply chain attacks, with an apparent increasing frequency~\cite{sonatype2026sscr,gokkaya2026software}. 
We did not observe any practitioners at either extreme end of the trust spectrum: no participant demonstrated full trust in the software supply chain by working without any controls, and no participant independently fully evaluated every dependency and every dependency update.
For example, participant \user{P23} acknowledged a \textit{verify-all zero-trust strategy} as a hypothetical option but considered it impractical: \qt{You can do it manually. You could go out and research each one of these components. That would take you a lot of time.}
Yet, practitioners' trust is evolving. 

\subsubsection*{\newfinding{finding:add-controls}{Practitioners add controls and rarely retire them, suggesting a decline in the trust extended to the supply chain.}}
Every participant described controls they adopt for their supply chain, and across all of them, the pattern is the same: Controls are added over time and almost never removed (unless replaced by another one; see Finding~\ref{finding:replace-controls}). The additions span all three areas we examined (dependencies, build infrastructure, and human factors) and include both manual and automated checks, with the newer additions being increasingly automated and AI-based controls.

Regarding dependency management, many participants added automated scanning for updates and known vulnerabilities in recent years, using \control{Software Composition Analysis (SCA) tools} like Snyk~\cite{snykopensource} and Dependabot~\cite{dependabot}  \user{P1, P2, P3, P6, P9, P10, P12, P13, P14, P17, P18, P19, P20, P22, P23, P24, P26, P28, P29, P34, P35, P36, P38}.
Several participants integrate \control{security gates} into their CI/CD pipelines to inspect dependencies and their updates (in addition to their own code) \user{P1, P2, P3, P10, P11, P14, P20, P22, P23, P24, P25, P31, P32, P34, P36}; while others use \control{LLMs} such as GitHub Copilot and Claude \user{P14, P18, P21, P24, P30, P32, P38}. 
Several participants also mentioned \control{automated SBOM generation} \user{P4, P10, P11, P16, P22, P23, P29}.
Specifically on updates, a few practitioners mentioned that they intentionally \control{defer updating dependencies} for at least a week or a month to avoid vulnerable updates \user{P11, P25, P28, P33}.
The same increase in controls appears in securing the build infrastructure.
Several participants (or their institution) adopt \control{reproducible builds} \user{P3, P6, P7, P20, P29, P34, P37}.
Several participants mentioned \control{reviewing third-party GitHub Actions} to avoid malicious actions \user{P1, P11, P21, P28, P29}.
Regarding third-party contributions, several participants mentioned performing \control{manual review} of either the code or the contributor \user{P4, P19, P29, P33, P34, P35}, using \control{CI/CD quality gates} \user{P4, P18, P19, P30, P34}, or both, to protect the project's integrity as an upstream artifact.

This accumulation of controls is described by practitioners as mostly reactive, as their awareness of software supply chain security increases over time \user{P8, P11, P20, P22, P23, P28, P29, P32, P37, P38}.
Practitioners mentioned external drivers, such as security incidents, including attacks and vulnerabilities, both those that happened to them directly and those they heard about in the news, and the influence of other people, like seniors. 
For example, \user{P28} mentioned \qt{the more I hear about [vulnerabilities news], the more I'm trying to \ldots provide more checks when necessary.}

Recall that we do not place a judgment on whether trust is appropriate. Practitioners may well argue that past levels of nearly blind trust are indefensible in the current environment with active attackers.
Increased controls may then reflect a better-calibrated level of appropriate (lower) trust, as mentioned by \user{P13}: \qt{[now] anyone can create [open-source packages, and] there could be [malicious] packages that maybe we shouldn't blindly install.}

\subsubsection*{\newfinding{finding:default}{Where risk and controls are not front of mind, trust is extended by default rather than chosen.}}
Trust is not always deliberately extended; Luhmann~\cite{luhmann1979trust} describes system trust as a way of reducing complexity, a means of not having to examine the world before acting, and running unexamined until something disrupts it. 
Much of the reliance we observed has this character: practitioners extend trust to the supply chain not after weighing the risks and the available controls, but due to intuitive judgments, community practices, or simply a lack of \textit{awareness} of risks and available options.
\looseness=-1

While all participants were aware of some risks, their awareness of their own exposure and of possible controls (e.g., \control{SCA tools}, \control{hermetic builds}) varied, generally with experience.  
Several practitioners mentioned that they became more aware of software supply chain security risks due to our interview request and some started considering controls as a reaction \user{P5, P7, P15, P19, P24, P28, P32}.
For example, \user{P24} realizes during the interview: \qt{Oh, I never actually thought... that even packages could be developed by LLMs. But now that I think about it...}

Unawareness was most visible among newcomers to open source, who do not think about security as they work, delegate dependency adoption and management, or build pipelines entirely to a coding agent; some self-describe as ``vibe coders.'' 
Their projects sometimes have some controls
in place, but most are through no decision of their own.
In industry, unawareness is sometimes deliberately cultivated among practitioners who can delegate trust decisions to a security team (Findings~\ref{finding:delegate-decision}--\ref{finding:delegate-control}).
Note that not adopting a control due to unawareness is different from a reasoned dismissal of the control  (e.g., 
\user{P3}'s \qt{the return on investment maybe isn't worth dedicating resources}). 
Both practitioners who deliberately forgo a control and those unaware of it still extend trust to the software supply chain, but only the former have chosen to do so.\looseness=-1

\subsubsection*{\newfinding{finding:cheap-signals}{Many control choices are still based on cheap, traditional signals, such as dependency popularity and maintainer activity, but AI puts pressure on these signals.}}

In line with past research~\cite{dabbish2012social,wermke2022committed,pashchenko2020qualitative, hamer2025trusting, bogart2016break, larios2020selecting, mujahid2023characteristics, qiu2019signals}, cheap, traditional signals about a component are still widely used to make decisions about adoption and updates, especially \control{popularity} \user{P2, P6, P11, P12, P17, P18, P19, P29, P37, P38}, \control{maintainer activity} and \control{documentation quality} \user{P1, P2, P3, P6, P7, P10, P11, P13, P14, P17, P19, P22, P28, P29, P33, P34, P38}, and \control{recommendations in communities} such as Stack Overflow and Reddit \user{P9, P12, P17, P32}.
A majority of participants said they would check how popular a dependency is, using it as a proxy for maintenance reliability and security/quality (the `many eyes` hypothesis~\cite{raymond1999cathedral}). 
We interpret the use of these signals rather than deeper checks as a sign of persisting trust in the software supply chain as a system~\cite{luhmann1979trust}.

With the rise of AI, several practitioners are using AI recommendations instead of the community's.
They ask the AI to recommend the best dependencies for their needs \user{P4, P5, P9, P18, P24, P32, P37, P38}, with at most some manual checks of popularity and activity signals \user{P4, P37}.
For example, \user{P37} mentioned \qt{despite referring to the LLM to make that judgment [of choosing dependencies], I would still probably [choose] the most popular for that specific use case.}

Only few participants \user{P14, P15, P19, P32} raised concerns about 
fakeability of these signals through coordinated malicious activities (fake stars~\cite{he2026six}) and AI; for example, \user{P19} considered anything below 1000 stars as unreliable because it is easy to fake. 
Several participants voiced concern about AI-generated content (Finding~\ref{finding:replace-controls}), but only \user{P32} connected it to the signals they depend on, no longer relying on documentation quality.
Generally, this issue was not at the forefront of our participants' minds, suggesting that AI had not yet broadly undermined system trust at the time of our interviews, though this may change.\looseness=-1




\subsubsection*{\newfinding{finding:personal}{Trust in people within the software supply chain stays personal: Practitioners rely on personal relationships and behavioral track records.}}

For software supply chain decisions that involve interpersonal interactions, such as assigning open-source maintainers, many personal trust signals based on past interactions are used and remain stable. 
For maintainers, several practitioners in collaborative projects report relying on personal trust relationships in decision-making: \control{in-person acquaintances} and a preference for \control{people they know} as maintainers \user{P4, P19, P28, P33, P34, P35, P37}, where they can judge their ability and integrity over time.
This reliance contrasts with decisions about code dependencies or builds, where most practitioners do not check dependency maintainers individually \user{P5, P9, P17, P21, P26, P30, P33, P34, P37, P38}.\looseness=-1
\subsection{Relocating Trust through Automation and Delegation}
\label{subsec:result_part2}

So far, we have observed a slow erosion of trust in the software supply chain through increasing adoption of controls. Now, we discuss strategies practitioners adopt to manage the corresponding costs, often by relocating trust elsewhere.



\subsubsection*{\newfinding{finding:replace-controls}{To reduce cost, manual controls are increasingly automated, which makes the automation a new object of trust.}}

Manual controls, such as \control{auditing dependency updates}, are expensive and scale poorly, and the reason participants cited most often for turning to \textit{automation} was exactly this failure to scale \user{P3, P7, P11, P12, P13, P18, P19, P23, P24, P30, P38}. 
Several described tools either use automation to guide where to spend manual effort, or partially or fully automate checks they previously performed by hand \user{P1, P2, P4, P10, P11, P12, P13, P16, P17, P18, P19, P20, P21, P23, P24, P25, P26, P27, P28, P30, P32, P33, P34, P35, P36, P37, P38}.
In dependency management, \control{automated scanning} replaces \control{manual inspection}, with manual attention reserved for what the tooling surfaces. 
In build infrastructure, automation replaces manual \control{inspection of logs and pipelines} \user{P2, P5, P17, P19, P22, P30}. 
When reviewing open-source contributions as upstream artifacts, two participants report no longer \control{reading the source code} directly but instead relying on \control{static analysis}, \control{CI/CD gates}, or an \control{AI reviewer} \user{P18, P19}.

We interpret this shift not as a matter of trusting the dependency more or less, but rather as a way to achieve a comparable level of control at lower cost. 
However, by automating the controls, the automation itself becomes a new object of trust.
Developers are exposed to risks if automation fails, particularly if it does so more often (or systematically) than the manual step it replaced. 
Many participants understand that these tools are not perfect and usually perform additional checks, either manually or by adding more tools, especially on dimensions they do not fully trust.
For example, several practitioners mentioned that they trust AI for speed and depth, but not for quality, which is the part where practitioners still check themselves \user{P14, P18, P21, P35, P38}.
Some practitioners intentionally use multiple tools/AIs to catch one another if one fails, akin to adversarial AI agents \user{P5, P15, P18, P20, P21, P30, P36}.
Not only do they trust AI only for some of its qualities, but practitioners also tend to trust AI when they use it themselves, but \textit{not} when other people use it (e.g., AI-generated dependencies) \user{P19, P27, P28, P32, P35, P37}.
Their trust in automation is therefore also conditioned on who controls it.\looseness=-1




\subsubsection*{\newfinding{finding:delegate-decision}{Some practitioners delegate decisions about trust: When a third party clears a component, they adopt it with little control of their own.}}



Rather than judging a component themselves, some practitioners hand the decision of \textit{what to trust} to a third party and rely on that party's verdict.
In the trust literature, this decision-maker is also known as a \textit{guardian of trust} \cite{shapiro1987social} -- an intermediary between the trustor and the trustee who is better placed to evaluate the trustee.
In the industry, the guardian is commonly the organization's security or compliance function. Several industry participants relied on such a team to supply a \control{dependency whitelist} \user{P3, P8, P14, P16, P22, P24, P25, P31}, often without being aware of what controls the guardian employs \user{P8, P14, P24, P31, P36}.
Several institutions even maintain an internal repository of vetted dependencies, so from the developers' perspective, they never adopt external dependencies \user{P8, P24, P31}.
The verdict can also be negative, as when \user{P10}'s organization was told to \qt{stop using GitHub Actions \ldots because our CISO did not trust GitHub Actions} -- again, a guardian made the decision for the developer.

Guardians are not always in-house. A few practitioners mentioned third-party guardians whose screening they trust, such as licensed open-source dependencies from Red Hat \user{P9, P23} and components available in specific marketplaces \user{P22}. 
Delegation is not always voluntary in corporations, though developers could pursue additional controls if they distrusted the guardian; none of our participants did. 

AI can effectively serve as a guardian too:
Some practitioners delegate the what-to-trust decision to an LLM \user{P18, P24, P32, P37}, as with \user{P37}: \qt{Claude chose that dependency by default for me.}
Here, AI is not used intentionally as a control to evaluate dependencies, but decision-making is fully delegated.

Delegating decisions requires developers to trust the guardian to do their job~\cite{shapiro1987social}, but it also outsources and pools the costs of controls, freeing developers' time and attention. 
For example, \user{P7} trusts the security team, expressing that they \qt{don't have the strength, nor the people who can sit and work on the security and understanding of each and every package [they use]}.
A guardian can also enact controls centrally for the developer where the developer would have just extended trust (deliberately or by default, cf. Findings~\ref{finding:default}, \ref{finding:cheap-signals}).
For example, \user{P14} mentioned that before joining the institution, they \qt{did not know enough} and \qt{probably would have just used any dependency,} yet after joining the institution, the institution's security team approves dependencies for them. 

\subsubsection*{\newfinding{finding:delegate-control}{Some practitioners delegate determining appropriate levels of control, accepting a checking regime that constrains how far practitioners may extend trust to the supply chain.}}




Unlike delegating decisions about what can be trusted, some practitioners report delegating decisions about controls to a guardian, who decides what level of trust is acceptable. 
Here, a third party like a central security team sets the required controls, for example, mandating \control{scans in the pipeline}, as mentioned by \user{P3}: \qt{tying the security scans into the CI/CD pipeline was a big component of what [the security team] has done. If you don't scan, you don't ship.} 
In addition to guardians within an organization like a security team, practitioners also delegate decisions about controls to recommendations of external guardians, including OpenSSF~\cite{openssf} \user{P3, P29}, OWASP~\cite{owasp} \user{P20, P23}, and NIST SSDF~\cite{nist_ssdf_2022} \user{P10}.

The delegated decision of controls to apply can be in the form of (i) the guardian telling developers which controls to use (recommended controls, suggested processes), or (ii) the guardian putting the controls in place as part of the infrastructure (forced process compliance, e.g., set up build infrastructure).
The effect is a \textit{cap on trust:} developers cannot extend trust beyond what the regime permits, even if they are inclined to do so. For example, if the regime mandates checking dependencies, developers cannot just extend trust and adopt an unchecked dependency even if they want to.
Whereas the delegation of decisions (Finding~\ref{finding:delegate-decision}) lets a developer do less, the delegation here requires them to do more, or at least bars them from doing less than the set level.

Developers' stance toward these mandated controls varies. 
Many accept it readily and are glad the decision is made for them by somebody with more expertise \user{P1, P7, P9, P13, P14, P16, P20, P24, P31, P37}, which is a sign \textit{system trust} extended to the institution, and a reduction of the complexity they would otherwise carry \cite{luhmann1979trust}.
Others experience the regime as imposed rather than chosen: Some developers accept the controls, but resent the costs \user{P14, P23, P25, P26} (e.g., \user{P25} acknowledges the necessity of the controls, yet \qt{as a developer, I would sometimes like to bypass security because I'm on tight deadlines. So it does get frustrating when all of these controls are enforced.}), and some consider the imposed controls as miscalibrated, overly conservative, or performative \user{P10, P11, P16, P23}.

\subsubsection*{\newfinding{finding:withdrawal}{Avoiding exposure entirely by withdrawing from the use of 
dependencies is rare, but raised as a possibility for an AI-assisted future.}}

A few participants pointed toward removing the exposure to a third party altogether rather than adopting controls \user{P2, P5, P12, P19, P21, P29, P34, P38}. 
\user{P12} argued that functionality once obtained from a third-party dependency could simply be regenerated: \qt{with AI, we can just rewrite the whole thing,} avoiding the exposure a dependency creates in the first place. 
This strategy may not scale beyond small dependencies; for example,  \user{P21} uses LLMs instead of adopting dependencies when dependencies have fewer than 1000 lines of code, but acknowledges they \qt{would never build like a Spring Framework from scratch}.
Beyond code dependencies, many third-party GitHub actions can also be replaced with AI-generated code \user{P12, P34, P32}.
This avoidance can be read as a severe mistrust of the supply chain, or simply as a realization that the production cost for some components has become lower than the transaction costs for control when adopting a third-party component in the absence of full trust. 
This exit strategy relocates trust to the automated code-generation process.


\subsection{Control Adoption to Build Trust}
\label{subsec:result_part3}

When talking to developers about why they adopt controls, it is usually clear that most controls are
adopted (or delegated) to reduce own risk exposure and can thus be read as an erosion of
trust. However, some participants also indicate that they adopt controls
to appear more trustworthy to others or because they are required by external parties. We read such controls as efforts to \textit{build} trust.



\subsubsection*{\newfinding{finding:voluntarily-gain-trust}{Some practitioners adopt controls as trust-building signals: Attestations, certifications, and policy adherence are adopted to appear trustworthy to downstream users.}}


Some practitioners are conscientious of their role as trustees to downstream users \user{P10, P13, P19, P20, P22, P29, P34}, and they adopt controls to signal to others. 
For example, \user{P34} performed reproducible builds and ensured full process transparency because that is \qt{needed to be trustable, not just by us, but by anybody else who came along and decided to use what we're using.}
%
Most controls intended as trust-building rather than for their own risk reduction relate to \control{attestations} \user{P19, P29, P34}, \control{software bills of materials (SBOM)} \user{P10, P22}, \control{certifications} \user{P13, P20}, and \control{documented policy adherence} \user{P10, P22}. 
They produce signals that are difficult to mimic, which is what makes them informative according to signaling theory (cf. Finding~\ref{finding:cheap-signals}).



\subsubsection*{\newfinding{finding:demanded-controls}{Some controls are a condition of doing business, demanded by clients and regulators.}}


Some controls are externally imposed \user{P1, P3, P4, P10, P11, P16, P22, P35, P36}, rather than the voluntary adoption to reduce risk exposure (Finding~\ref{finding:add-controls}) or the voluntary adoption to signal trustworthiness (Finding~\ref{finding:voluntarily-gain-trust}). For example, some clients (including government clients) may request a software bill of materials (SBOM) with no known vulnerabilities in the dependency tree. In interviews, the distinction
is clear when asking about reasons for controls, and we do not consider imposed controls as a lack of trust
of the developer, but as a sign of the trust posture of the party demanding them and of the larger system -- that reliance now has to be underwritten by an explicit external requirement (Zucker~\cite{zucker1986production}'s institution-based trust).

Imposed controls are often a \textit{client requirement} \user{P1, P10, P11, P16, P22, P35}, such as \user{P22} bringing in legal review when a change in investors made it a condition of that relationship: \qt{We started getting more into the licensing component of it and bringing in legal because we're changing investors.}
There are also several market contractual standards, such as PCI DSS~\cite{pcidss_v4}, that institutions must comply with to enter into a contract \user{P11, P22}. 
Several clients also requested compliance with SOC 2~\cite{soc2} \user{P10, P35} or ISO 27001~\cite{iso27001} \user{P1, P10, P16}. 
Governments can require specific controls through \textit{regulation} (participants mention EO-14028 \user{P3, P10} and GDPR \user{P16}), especially if they are customers. For example, \user{P3} reports \qt{doing attestations, \ldots basically since [EO-14028] \ldots if you want government contracts, yeah.} 

Big clients and governments can set expectations beyond their direct contracts, so that their imposed controls become a template adopted by others, even where no regulation binds them. 
This pattern is the bridge to voluntary signaling (Finding~\ref{finding:voluntarily-gain-trust}), where a control required of some becomes a signal volunteered by others \user{P10, P11, P30}.
For example, \user{P10}'s institution is not legally bound to EO-14028, yet they implemented a version of NIST SSDF~\cite{nist_ssdf_2022} that is imposed under EO-14028 for federal contractors.



\section{Discussions, Implications, and Conclusion}
\label{sec:discussion}
In this section, we situate our findings in broader trust literature, derive implications for key stakeholders, and suggest that the goal is not to restore blind trust but to calibrate an appropriate level of system trust.

\subsection{Reading Trust from Revealed Preferences}

We argued that trust in the supply chain is best understood through practitioners' actions, and that an explicit theoretical lens turns those actions into a coherent account.
This approach aligns with how trust is conceptualized as a construct: Luhmann~\cite{luhmann1979trust} describes system trust as running unexamined until something disrupts it, a way of not having to inspect the world before acting. 
Trust of this kind is, by its nature, not available to introspection until violated, whereas practices can provide insights.
Finding~\ref{finding:default} illustrates this: practitioners who adopt dependencies with no controls of their own hold a trust posture unlikely to surface in a survey, since their trust shows in \textit{not} thinking to check.
When participants talked about perceptions (unprompted or at the very end), these broadly tracked their actions, though not always.
For exampleFinding, \user{P35} called supply chain security \qt{getting better} while having \qt{removed [third-party GitHub actions] over time}; \user{P14} called dependency security \qt{getting worse} yet added no controls, delegating to a security team instead.\looseness=-1

The theory lens also gives us vocabulary to study common phenomena across participants, such as guardians, and clarifies friction points: the same control action could be read as mistrust or as trust-building, depending on the reason for adopting it.
That adoption reason is not observable in the action alone and is what the interview and lens together recover (Findings~\ref{finding:voluntarily-gain-trust}--\ref{finding:demanded-controls}).

\noindent \textbf{Implications for researchers.}
We found both the revealed-preferences approach and the theory lens valuable, and encourage researchers studying similarly slippery concepts to consider a comparable design. The two reinforce each other: Revealed preferences surface what practitioners do, and the theory supplies the concepts to interpret it (system trust, guardians, signals).

\subsection{From Interpersonal to System Trust}
We started our exploration of trust with Mayer's framing of \textit{interpersonal} trust~\cite{mayer1995integrative}, expecting to map actions to judgments about a maintainer's ability, integrity, benevolence, and propensity to trust.
This framing captured too small a slice:
almost all reliance in open source is on artifacts and large, loosely connected maintainer groups, not the small circles of acquaintance where personal trust operates (exceptions in Finding~\ref{finding:personal}).
The dominant phenomenon is \textit{system trust}~\cite{luhmann1979trust}, confidence in the supply chain as an arrangement rather than in any individual within it, the `lubricant` that enables cheap reuse in Arrow's terminology~\cite{arrow1974limits}. 
This perspective matches the observed erosion of trust without anyone becoming less trustworthy individually: \textit{system} trust is becoming more expensive to extend, rather than interpersonal trust breaking down.\looseness=-1

\noindent\textbf{Implications for practitioners and regulators.}
A system-trust perspective guides where interventions are more promising: actions that lift the system, not judgments of individuals.
For example, making transparent which maintainers enable two-factor authentication (2FA) might build personal trust. But enforcing 2FA across PyPI or npm hardens the entire system, making it harder to turn a benign maintainer into an attack vector and enabling high-trust interactions broadly.
The same logic applies to signals: as cheap signals degrade and are easily faked (Finding~\ref{finding:cheap-signals}), investing in harder-to-fake signals (e.g., provenance, attestations) not only lowers individual risk but raises the floor for everyone downstream (Findings~\ref{finding:voluntarily-gain-trust}--\ref{finding:demanded-controls}).
Guardians who vet once for many (Findings~\ref{finding:delegate-decision}--\ref{finding:delegate-control}) and the standards behind them (Finding~\ref{finding:demanded-controls}) work the same way: raising the floor for everyone rather than redistributing the cost of suspicion.
The most effective interventions strengthen system-level trust (standards and shared mechanisms that ``lift all boats''), rather than intensifying per-actor checks, which raise costs without rebuilding the regime.

\subsection{Learning from Other Disciplines: Low Trust Environment}

A low-trust environment limits cooperation and requires greater verification~\cite{fukuyama1995trust}.
In organizational settings, low trust correlates with low productivity and engagement~\cite{zak2017neuroscience}.
Equivalently, in our interviews, practitioners who trust dependencies are more willing to engage with them (\qt{I feel okay installing many packages.}\user{P13}), while those who have lost trust withdraw (\qt{I stopped relying totally on packages.}\user{P12})
In economics, low trust raises transaction costs~\cite{denbutter2003tradetrust}, which is equivalent to the rising cost of verification in software supply chain security.
In politics, low trust reduces institutional performance~\cite{putnam1993making, fukuyama1995trust}.
Analogously, developers who distrust their security team will double-check the team's decisions, extending production time and reducing output.


The natural response to a low-trust environment is delegation (Findings~\ref{finding:delegate-decision}--\ref{finding:delegate-control}); not indifference, but pooling checking effort to others with more expertise to reduce the transaction cost of repeated independent verification~\cite{arrow1974limits}. 
A guardian who vets once for many absorbs that redundant effort, though this requires trusting the guardian's competence and risks free-riding.
Delegating to AI is similar but without the pooling: an LLM is not a shared guardian vetting once for many, and it carries no accountability (the tension named by~\user{P4, P13, P16, P34, P38}).
We therefore believe external guardians who vet once for many are the most promising direction (Findings~\ref{finding:delegate-decision}--\ref{finding:delegate-control}).
Other disciplines, such as food safety and financial auditing, have constructed accredited institutions as guardians for consumers to delegate their trust to.
Early equivalents are emerging in the software supply chain; one was even mentioned by some participants: Mozilla's trusted open-source initiative~\cite{mozilla_moss}, Google's Assured OSS program~\cite{google_assured_oss}, Red Hat's licensed and enterprise-supported open-source distributions~\cite{redhat_open_source_assurance} \user{P9, P23}, and Chainguard's~\cite{chainguard_trusted_oss}. 

Guardians come with known limitations. 
Relocating trust to guardians concentrates rather than dissolves the trust problem~\cite{shapiro1987social}: the guardian becomes a high-value target for attackers and must be held to a higher standard than the components it clears. 
Guardians can also entrench incumbents, as established packages already have a track record while newer ones may struggle to gain entry, possibly undercutting innovation in the ecosystem. 
These limitations call for oversight.

\noindent \textbf{Implication for policy makers.}
Regulators can implement more systemic interventions by incentivizing, regulating, and certifying the guardian institutions that absorb verification costs (Findings~\ref{finding:delegate-decision}--\ref{finding:delegate-control}), holding them to the high standard their role demands.
To avoid systemic failure, policy could establish and fund verification bodies, certify guardians, and ensure sufficient redundancy.

\subsection{Calibrating Trust in Software Supply Chain}

When personal trust fails, repair strategies such as apologies and accountability mechanisms restore the prior state~\cite{dirks2009reparing}.
Yet, in the software supply chain, practitioners extend system trust: there is no single transgression to repair and no party to hold accountable, but a gradual erosion that makes trust more expensive.
The goal is therefore not restoration but \textit{calibration}: aligning trust with actual trustworthiness~\cite{lee2004trust}.
Miscalibrations in either direction are costly: overtrust (e.g., old, near-blind trust in an environment with capable adversaries) carries hidden risk, and undertrust is costly due to excessive controls that exceed the risks they address.
Other domains calibrate through risk-tiered verification; food safety inspectors, for instance, scrutinize high-risk ingredients more closely. 
The software supply chain equivalent is matching the control level to the dependency criticality
and to signal quality (e.g., attestation is harder for AI to mimic and warrants more trust than popularity alone).\looseness=-1

\noindent\textbf{Implications for security leaders.} 
With attacks now constant rather than episodic, calibration provides security leaders and guardians with a principled basis to defend to skeptical developers and budget holders (Finding~\ref{finding:delegate-control}), rather than ratcheting up controls after every incident. 
This shift enables honest dialogue about risk instead of reactive justification.

\subsection{Return to the Opening Question}
We opened with the question \textit{``Are practitioners losing trust in the software supply chain?''} Our study lends support to the common perception that trust is eroding, doing so through the lens of costly actions practitioners actually take rather than through stated opinions. However, our study presents a more nuanced and far less bleak picture than some voices suggest.

An intuitive \textit{``well, yes, obviously it is getting worse''}, flattens nuances of high personal trust and eroding system trust, trust relocation toward automation (e.g., trusting their own use of AI, but not AI use of others), guardians who vet once for many, and external trust-building forces. 
Trust is eroding, yes, but the software supply chain is not collapsing into a low-trust environment with excessive transaction costs (the equivalent of a society where everyone withdraws to close personal ties and hires security guards).
Instead, a high-trust regime is reorganizing (``relubricating'') itself under adversarial pressure. These trust mechanisms, not the trend direction, are what future tools, policies, and governance can act on.

\bibliographystyle{IEEEtran}
\bibliography{short,bibfile}

@String{CACM = "Communications of the {ACM}" }

@String{CACM = "Commun. {ACM}" }

@String{Academic = "Academic Press" }

@String{Springer = "Springer-Verlag" }

@article{zucker1986production,
  title={Production of trust: Institutional sources of economic structure, 1840--1920.},
  author={Zucker, Lynne G},
  journal=ROB,
  year={1986},
  publisher={JAI Press, Inc.}
}

@article{mayer1995integrative,
  title={An integrative model of organizational trust},
  author={Mayer, Roger C and Davis, James H and Schoorman, F David},
  journal=AMR,
  volume={20},
  number={3},
  pages={709--734},
  year={1995},
  publisher={Academy of Management Briarcliff Manor, NY 10510}
}

@misc{dependabot,
  author = {{GitHub}},
  title = {Dependabot},
  howpublished = {\href{https://github.com/dependabot}{Dependabot}},
  year = {2026},
  note = {Automated dependency management}
}

@inproceedings{bogart2016break,
  title={How to break an API: cost negotiation and community values in three software ecosystems},
  author={Bogart, Christopher and K{\"a}stner, Christian and Herbsleb, James and others}, 
  booktitle=FSE-16,
  pages={109--120},
  year={2016}
}

@inproceedings{pashchenko2020qualitative,
  title={A qualitative study of dependency management and its security implications},
  author={Pashchenko, Ivan and Vu, Duc-Ly and Massacci, Fabio},
  booktitle=CCS-20,
  pages={1513--1531},
  year={2020}
}

@inproceedings{he2026six,
  title={Six Million (Suspected) Fake Stars on GitHub: A Growing Spiral of Popularity Contests, Spam, and Malware},
  author={He, Hao and Yang, Haoqin and Burckhardt, Philipp and others}, 
  booktitle=ICSE-26,
  year={2026}
}

@book{luhmann1979trust,
  title     = {Trust and Power},
  author    = {Luhmann, Niklas},
  year      = {1979},
  publisher = {John Wiley \& Sons},
  address   = {Chichester, England}
}

@book{fukuyama1995trust,
  author    = {Francis Fukuyama},
  title     = {Trust: The Social Virtues and the Creation of Prosperity},
  publisher = {Free Press},
  address   = {New York},
  year      = {1995},
  isbn      = {9780029109762}
}

@book{putnam1993making,
  author    = {Robert D. Putnam},
  title     = {Making Democracy Work: Civic Traditions in Modern Italy},
  publisher = {Princeton University Press},
  address   = {Princeton, NJ},
  year      = {1993},
  isbn      = {9780691037387}
}

@article{zak2017neuroscience,
  title={The neuroscience of trust},
  author={Zak, Paul J},
  journal=HBR,
  volume={95},
  number={1},
  pages={84--90},
  year={2017}
}

@misc{denbutter2003tradetrust,
  author       = {Frank A. G. den Butter and Robert H. J. Mosch},
  title        = {Trade, Trust and Transaction Costs},
  year         = {2003},
  month        = oct,
  howpublished = {Tinbergen Institute Working Paper No. 2003-082/3},
  doi          = {10.2139/ssrn.459501},
  note         = {Available at SSRN}
}

@article{dirks2009reparing,
  title= {Reparing relationships within and between organizations: building a conceptual foundation},
  author={Dirks, Kurt T and Lewicki, Roy J and Zaheer, Akbar},
  journal=AMR,
  volume={34},
  number={1},
  pages={68--84},
  year={2009},
  publisher={Academy of Management Briarcliff Manor, NY}
}

@inproceedings{boughton2024decomposing,
  title={Decomposing and measuring trust in open-source software supply chains},
  author={Boughton, Lina and Miller, Courtney and Acar, Yasemin and others}, 
  booktitle=ICSE-NIER-24,
  pages={57--61},
  year={2024}
}

@article{hou2023systematic,
  title={A systematic literature review on trust in the software ecosystem},
  author={Hou, Fang and Jansen, Slinger},
  journal=ESEJ,
  volume={28},
  number={1},
  pages={8},
  year={2023},
  publisher={Springer}
}

@techreport{bauer2019conceptualizing,
  title={Conceptualizing trust and trustworthiness},
  author={Bauer, Paul C},
  year={2019},
  institution={Revised version of working paper published in: Political Concepts Working Paper Series}
}

@inproceedings{kula2015trusting,
  title={Trusting a library: A study of the latency to adopt the latest maven release},
  author={Kula, Raula Gaikovina and German, Daniel M and Ishio, Takashi and others},
  booktitle=SANER-15,
  pages={520--524},
  year={2015},
  organization={IEEE}
}

@inproceedings{sammak2023developers,
  title={Developers' approaches to software supply chain security: An interview study},
  author={Sammak, Rami and Rotthaler, Anna Lena and Ramulu, Harshini Sri and others}, 
  booktitle=SCORED-24,
  pages={56--66},
  year={2023}
}

@inproceedings{wermke2022committed,
  title={Committed to trust: A qualitative study on security \& trust in open source software projects},
  author={Wermke, Dominik and W{\"o}hler, Noah and Klemmer, Jan H and others}, 
  booktitle=SNP-22,
  pages={1880--1896},
  year={2022},
  organization={IEEE}
}

@article{jansen2021trustseco,
  title={TrustSECO: an interview survey into software trust},
  author={Jansen, Floris and Jansen, Slinger and Hou, Fang},
  journal={arXiv preprint arXiv:2101.06138},
  year={2021}
}

@book{arrow1974limits,
  title={The limits of organization},
  author={Arrow, Kenneth J},
  year={1974},
  publisher={WW Norton \& Company}
}

@book{putnam2000bowling,
  title={Bowling alone: The collapse and revival of American community},
  author={Putnam, Robert D},
  year={2000},
  publisher={Simon and schuster}
}

@article{Spence1973,
  author  = {Spence, Michael},
  title   = {Job Market Signaling},
  journal = QJE,
  volume  = {87},
  number  = {3},
  pages   = {355--374},
  year    = {1973},
  doi     = {10.2307/1882010}
}

@article{qiu2019signals, title={The signals that potential contributors look for when choosing open-source projects}, 
author={Qiu, Huilian Sophie and Li, Yucen Lily and Padala, Susmita and others},  
journal=PACM-HCI-19, volume={3}, number={CSCW}, pages={1--29}, year={2019}, publisher={ACM New York, NY, USA} }

@article{hamer2025trusting,
  title={Trusting code in the wild: Exploring contributor reputation measures to review dependencies in the Rust ecosystem},
  author={Hamer, Sivana and Imtiaz, Nasif and Tamanna, Mahzabin and others},
  journal=TSE,
  year={2025},
  publisher={IEEE}
}

@misc{freund2024xz,
  author = {Andres Freund},
  title = {Backdoor in Upstream XZ/liblzma Leading to SSH Server Compromise},
  year = {2024},
  howpublished = {Openwall OSS Security Mailing List},
  month = mar,
  note = {Posted March 29, 2024. \href{https://www.openwall.com/lists/oss-security/2024/03/29/4}{openwall.com}},
}

@article{roose2024oneguy,
  author = {Kevin Roose},
  title = {Did One Guy Just Stop a Huge Cyberattack?},
  journal = {The New York Times},
  year = {2024},
  month = apr,
  day = {3},
  howpublished = {\href{https://www.nytimes.com/2024/04/03/technology/prevent-cyberattack-linux.html}{nytimes.com/prevent-cyberattack}}
}

@inproceedings{williams2025can,
  title={Can the Rising Tide of Software Supply Chain Attacks Raise All Software Engineering Boats?},
  author={Williams, Laurie and Hamer, Sivana and Zahan, Nusrat},
  booktitle=FSE-25,
  pages={18--26},
  year={2025}
}

@article{williamsDirection2025,
author={Williams, Laurie and Benedetti, Giacomo and Hamer, Sivana and others},
title = {Research Directions in Software Supply Chain Security},
year = {2025},
issue_date = {June 2025},
publisher = {ACM},
address = {New York, NY, USA},
volume = {34},
number = {5},
issn = {1049-331X},
doi = {10.1145/3714464},
journal = {ACM Trans. Softw. Eng. Methodol.},
articleno = {146},
numpages = {38}
}

@techreport{sonatype2026sscr,
  author       = {{Sonatype}},
  title        = {2026 State of the Software Supply Chain Report},
  institution  = {Sonatype},
  year         = {2026},
  month        = jan,
  howpublished          = {\href{https://www.sonatype.com/hubfs/1-2025_Website-Assets/SSCR\_2025/SSCR\_2026\_final.pdf}{SSCR\_2026\_final.pdf}},
  note         = {Accessed: 2026-06-16}
}

@article{gokkaya2026software,
  title={Software supply chain: A taxonomy of attacks, mitigations and risk assessment strategies},
  author={Gokkaya, Betul and Aniello, Leonardo and Halak, Basel},
  journal=JISA,
  volume={97},
  pages={104324},
  year={2026},
  publisher={Elsevier}
}

@article{raymond1999cathedral,
  title={The cathedral and the bazaar},
  author={Raymond, Eric},
  journal=KTP,
  volume={12},
  number={3},
  pages={23--49},
  year={1999},
  publisher={Springer}
}

@article{de2010can,
  title={How can contributors to open-source communities be trusted? On the assumption, inference, and substitution of trust},
  author={De Laat, Paul B},
  journal=EIT,
  volume={12},
  number={4},
  pages={327--341},
  year={2010},
  publisher={Springer}
}

@article{samuelson1938,
  author    = {Paul A. Samuelson},
  title     = {A Note on the Pure Theory of Consumers' Behaviour},
  journal   = {Economica},
  series    = {New Series},
  volume     = {5},
  number     = {17},
  pages      = {61--71},
  year       = {1938},
  doi        = {10.2307/2548836},
  jstor      = {2548836}
}

@incollection{paulhus1991measurement,
  author    = {Paulhus, Delroy L.},
  title     = {Measurement and Control of Response Bias},
  booktitle = {Measures of Personality and Social Psychological Attitudes},
  editor    = {Robinson, John P. and Shaver, Phillip R. and Wrightsman, Lawrence S.},
  pages     = {17--59},
  publisher = {Academic Press},
  address   = {San Diego, CA},
  year      = {1991},
}

@article{glaeser2000measuring,
  title={Measuring trust},
  author={Glaeser, Edward L and Laibson, David I and Scheinkman, Jose A and others}, 
  journal=QJE,
  volume={115},
  number={3},
  pages={811--846},
  year={2000},
  publisher={MIT Press}
}

@article{fereday2006demonstrating,
  title={Demonstrating rigor using thematic analysis: A hybrid approach of inductive and deductive coding and theme development},
  author={Fereday, Jennifer and Muir-Cochrane, Eimear},
  journal=IJQM,
  volume={5},
  number={1},
  pages={80--92},
  year={2006},
  publisher={Sage Publications CA}
}

@article{braun2006using,
  title={Using thematic analysis in psychology},
  author={Braun, Virginia and Clarke, Victoria},
  journal=QRP,
  volume={3},
  number={2},
  pages={77--101},
  year={2006},
  publisher={Taylor \& Francis}
}

@book{lincoln1985naturalistic,
  title={Naturalistic inquiry},
  author={Lincoln, Yvonna S},
  volume={75},
  year={1985},
  publisher={sage}
}

@article{parnas1972criteria,
  title={On the criteria to be used in decomposing systems into modules},
  author={Parnas, David Lorge},
  journal=CACM,
  volume={15},
  number={12},
  pages={1053--1058},
  year={1972},
  publisher={ACm New York, NY, USA}
}

@article{soto2021comprehensive,
  title={A comprehensive study of bloated dependencies in the maven ecosystem},
  author={Soto-Valero, C{\'e}sar and Harrand, Nicolas and Monperrus, Martin and others}, 
  journal=ESEJ,
  volume={26},
  number={3},
  pages={45},
  year={2021},
  publisher={Springer}
}

@article{decan2019empirical,
  title={An empirical comparison of dependency network evolution in seven software packaging ecosystems},
  author={Decan, Alexandre and Mens, Tom and Grosjean, Philippe},
  journal=ESEJ,
  volume={24},
  number={1},
  pages={381--416},
  year={2019},
  publisher={Springer}
}

@inproceedings{necula1997proof,
  title={Proof-carrying code},
  author={Necula, George C},
  booktitle=POPL-97,
  pages={106--119},
  year={1997}
}

@inproceedings{dabbish2012social,
  title={Social coding in GitHub: transparency and collaboration in an open software repository},
  author={Dabbish, Laura and Stuart, Colleen and Tsay, Jason and Herbsleb, Jim},
  booktitle=CSCW-12,
  pages={1277--1286},
  year={2012}
}

@article{DasTeng1998,
  author  = {Das, T. K. and Teng, Bing-Sheng},
  title   = {Between Trust and Control: Developing Confidence in Partner Cooperation in Alliances},
  journal = AMR,
  volume  = {23},
  number  = {3},
  pages   = {491--512},
  year    = {1998},
  doi     = {10.5465/amr.1998.926623}
}

@article{PoppoZenger2002,
  author  = {Poppo, Laura and Zenger, Todd},
  title   = {Do Formal Contracts and Relational Governance Function as Substitutes or Complements?},
  journal = SMJ,
  volume  = {23},
  number  = {8},
  pages   = {707--725},
  year    = {2002},
  doi     = {10.1002/smj.249}
}

@article{akerlof1970market,
  author    = {Akerlof, George A.},
  title     = {The Market for ``Lemons'': Quality Uncertainty and the Market Mechanism},
  journal   = QJE,
  volume    = {84},
  number    = {3},
  pages     = {488--500},
  year      = {1970},
  publisher = {MIT Press}
}

@article{finken2025characteristics,
  author  = {Tanner Finken and Jesse Chen and Sazzadur Rahaman},
  title   = {On the Characteristics and Impacts of Protestware Libraries},
  journal = PACMSE-25,
  volume  = {2},
  number  = {FSE},
  pages   = {2500--2523},
  year    = {2025},
  publisher = {ACM},
  doi     = {10.1145/3729381}
}

@book{Perrow1984,
  author    = {Charles Perrow},
  title     = {Normal Accidents: Living with High-Risk Technologies},
  publisher = {Basic Books},
  address   = {New York},
  year      = {1984}
}

@article{McKnight01122001,
author = {D. Harrison McKnight and Norman L. Chervany},
title = {What Trust Means in E-Commerce Customer Relationships: An Interdisciplinary Conceptual Typology},
journal = IJEC,
volume = {6},
number = {2},
pages = {35--59},
year = {2001},
publisher = {Routledge},
doi = {10.1080/10864415.2001.11044235},
}

@book{searcy2005evolution,
  title={The Evolution of Animal Communication: Reliability and Deception in Signaling Systems},
  author={Searcy, William A and Nowicki, Stephen},
  year={2005},
  publisher={Princeton University Press}
}

@article{shapiro1987social,
  title={The social control of impersonal trust},
  author={Shapiro, Susan P},
  journal=AJS,
  volume={93},
  number={3},
  pages={623--658},
  year={1987},
  publisher={University of Chicago Press}
}

@ARTICLE{Zahan2024-vz,
  title         = "{S3C2} summit 2023-11: Industry Secure Supply Chain Summit",
  author = "Zahan, Nusrat and Acar, Yasemin and Cukier, Michel and others",        
  journal       = "arXiv preprint arXiv:2408.16529",
  year          =  2024,
  archivePrefix = "arXiv",
  primaryClass  = "cs.CR"
}

@article{jowsey2025we,
  title={We reject the use of generative artificial intelligence for reflexive qualitative research},
  author={Jowsey, Tanisha and Braun, Virginia and Clarke, Victoria and others}, 
  journal=QI,
  pages={10778004251401851},
  year={2025},
  publisher={Sage Publications CA}
}

@misc{snykopensource,
  author       = {{Snyk Ltd.}},
  title        = {Snyk Open Source},
  year         = {2026},
  howpublished  = {\href{https://snyk.io/product/open-source-security-management/}{snyk.io}},
  urldate      = {2026-06-24},
  note         = {Software Composition Analysis (SCA) tool}
}

@techreport{nist_ssdf_2022,
  author      = {Souppaya, Murugiah and Scarfone, Karen and Dodson, Donna},
  title       = {{Secure Software Development Framework (SSDF) Version 1.1: Recommendations for Mitigating the Risk of Software Vulnerabilities}},
  institution = {National Institute of Standards and Technology},
  year        = {2022},
  number      = {NIST Special Publication 800-218},
  doi         = {10.6028/NIST.SP.800-218},
}

@misc{openssf,
  author       = {{Open Source Security Foundation}},
  title        = {{Open Source Security Foundation (OpenSSF)}},
  year         = {2026},
  howpublished = {\href{https://openssf.org}{openssf.org}},
  note         = {Accessed: 2026-06-24}
}

@misc{owasp,
  author       = {{OWASP Foundation}},
  title        = {{OWASP Foundation}},
  year         = {2026},
  howpublished = {\href{https://owasp.org}{owasp.org}},
  note         = {Accessed: 2026-06-24}
}

@misc{soc2,
  author       = {{American Institute of Certified Public Accountants}},
  title        = {{SOC for Service Organizations: Trust Services Criteria}},
  year         = {2022},
  howpublished = {\href{https://www.aicpa-cima.com}{aicpa-cima.com}},
  note         = {Accessed: 2026-06-25}
}

@misc{pcidss_v4,
  author       = {{PCI Security Standards Council}},
  title        = {{Payment Card Industry Data Security Standard (PCI DSS) Version 4.0.1}},
  year         = {2024},
  howpublished = {\href{https://www.pcisecuritystandards.org}{pcisecuritystandards.org}},
  note         = {Accessed: 2026-06-25}
}

@techreport{iso27001,
  author      = {{ISO/IEC}},
  title       = {{ISO/IEC 27001:2022 Information Security, Cybersecurity and Privacy Protection}},
  institution = {{International Organization for Standardization} and {International Electrotechnical Commission}},
  year        = {2022},
  number      = {ISO/IEC 27001:2022},
  address     = {Geneva, Switzerland},
  howpublished = {\href{https://www.iso.org/standard/27001}{iso.org}}
}

@misc{mozilla_moss,
  author = {{Mozilla Foundation}},
  title  = {{Mozilla Open Source Support (MOSS)}},
  howpublished    = {\href{https://www.mozilla.org/en-US/moss/}{mozilla.org}},
  note   = {Accessed: June 25, 2026}
}

@misc{google_assured_oss,
  author = {{Google Cloud}},
  title  = {{Assured Open Source Software}},
  howpublished    = {\href{https://cloud.google.com/assured-open-source-software}{cloud.google.com}},
  note   = {Accessed: June 25, 2026}
}

@misc{chainguard_trusted_oss,
  author = {{Chainguard}},
  title  = {{The Trusted Source for Open Source}},
  howpublished    = {\href{https://www.chainguard.dev/open-source}{chainguard.dev}},
  note   = {Accessed: June 25, 2026}
}

@misc{redhat_open_source_assurance,
  author = {{Red Hat}},
  title  = {{Open Source Assurance}},
  howpublished    = {\href{https://www.redhat.com/en/about/open-source-assurance}{redhat.com}},
  note   = {Accessed: June 25, 2026}
}

@misc{miro,
  author       = {{Miro}},
  title        = {{Miro: The Innovation Workspace}},
  year         = {2026},
  howpublished = {\href{https://miro.com/}{miro.com}},
  note         = {Online collaborative whiteboard platform.}
}

@techreport{ji2024cybersecurity,
  author      = {Ji, Jessica and Jun, Jenny and Wu, Maggie and others}, 
  title       = {Cybersecurity Risks of {AI}-Generated Code},
  institution = {Center for Security and Emerging Technology},
  year        = {2024},
  howpublished = {\href{https://cset.georgetown.edu/wp-content/uploads/CSET-Cybersecurity-Risks-of-AI-Generated-Code.pdf}{CSET.pdf}}
}

@inproceedings{larios2020selecting,
  title={Selecting third-party libraries: The practitioners’ perspective},
  author={Larios Vargas, Enrique and Aniche, Maur{\'\i}cio and Treude, Christoph and others},
  booktitle=ESEC-FSE-20,
  pages={245--256},
  year={2020}
}

@article{mujahid2023characteristics,
  title={What are the characteristics of highly-selected packages? A case study on the npm ecosystem},
  author={Mujahid, Suhaib and Abdalkareem, Rabe and Shihab, Emad},
  journal=JSS,
  volume={198},
  pages={111588},
  year={2023},
  publisher={Elsevier}
}

@article{lee2004trust,
  title={Trust in automation: Designing for appropriate reliance},
  author={Lee, John D and See, Katrina A},
  journal=HF,
  volume={46},
  number={1},
  pages={50--80},
  year={2004},
  publisher={SAGE Publications Sage UK: London, England}
}

@misc{zenodo,
  author       = {Anonymous},
  title        = {Supplementary Materials for Relubricating the Software Supply Chain: How Practitioners’ Trust Evolves Through Automation and Guardians},
  year         = 2026,
  publisher    = {Zenodo},
  doi          = {10.5281/zenodo.21077269},
  howpublished = {\href{https://zenodo.org/records/21077269?preview=1&token=eyJhbGciOiJIUzUxMiIsImlhdCI6MTc4Mjg0OTAxOCwiZXhwIjoxOTA5MDA3OTk5fQ.eyJpZCI6IjVjOWQ4ZTk3LTE4NTMtNGJlNS1hODc1LTI1ODJjMWMzNDQzZiIsImRhdGEiOnt9LCJyYW5kb20iOiJiMTA3MTNjM2ViMWU4MWYyNmE0YWQ4ZmViMjkwMzA1NSJ9.3imbofEVuHAO-sGzsFfJ9rco6QCuw8pf-cQ7hzT5AcwfzVWxVtMa94mTTTAJPsa7492jdqz87TOijtmH2GA9Hg}{Zenodo link}}
}

@article{malkawi2026ai,
  title={AI-Powered Vulnerability Detection and Patch Management in Cybersecurity: A Systematic Review of Techniques, Challenges, and Emerging Trends},
  author={Malkawi, Malek and Alhajj, Reda},
  journal=MAKE,
  volume={8},
  number={1},
  pages={19},
  year={2026},
  publisher={MDPI}
}

@string{ICSE-26 = "Proc.\ IEEE/ACM ICSE'26"}

@string{ICSE-NIER-24 = "Proc.\ IEEE/ACM ICSE-NIER'24"}

@string{FSE-16 = "Proc.\ ACM SIGSOFT FSE'16"}

@string{FSE-25 = "Proc.\ ACM FSE'25"}

@string{ESEC-FSE-20 = "Proc.\ ACM ESEC/FSE'20"}

@string{CCS-20 = "Proc.\ ACM SIGSAC CCS'20"}

@string{SANER-15 = "Proc.\ IEEE SANER'15"}

@string{SCORED-24 = "Proc.\ ACM SCORED'24"}

@string{SNP-22 = "Proc.\ IEEE S\&P'22"}

@string{POPL-97 = "Proc.\ ACM SIGPLAN-SIGACT POPL'97"}

@string{PACM-HCI-19 = "Proc. ACM HCI'19"}

@string{CSCW-12 = "Proc.\ ACM CSCW'12"}

@string{PACMSE-25 = "Proc. ACM Softw. Eng.'25"}

@string{ROB = "Res. Organ. Behav."}

@string{AMR = "Acad. Manage. Rev."}

@string{AS = "Adm. Soc."}

@string{HBR = "Harv. Bus. Rev."}

@String{ESEJ = "Emp. SW Eng."}

@string{JSS = "J. Syst. Softw."}

@string{HF = "Hum. Factors"}

@string{QJE = "Q. J. Econ."}

@string{TSE = "IEEE Trans. Softw. Eng."}

@string{JISA = "J. Inf. Secur. Appl."}

@string{KTP = "Knowl. Technol. Policy"}

@string{EIT = "Ethics Inf. Technol."}

@string{IJQM = "Int. J. Qual. Methods"}

@string{QRP = "Qual. Res. Psychol."}

@string{SMJ = "Strateg. Manage. J."}

@string{IJEC = "Int. J. Electron. Commer."}

@string{AJS = "Am. J. Sociol."}

@string{QI = "Qual. Inq."}

@string{MAKE = "Mach. Learn. Knowl. Extr."}

\end{document}